\documentclass[aip,jap,twocolumn,superscriptaddress,numerical,reprint]{revtex4-1}

\usepackage[utf8]{inputenc}

\usepackage{graphicx}
\usepackage{natbib}
\usepackage{color}
\usepackage[normalem]{ulem}

\newcommand{\tis}{TiS$_2$}
\newcommand{\tixs}[1]{Ti$_{#1}$S$_2$}
\newcommand{\tatis}{Ta$_{0.1}$Ti$_{0.9}$S$_2$}
\newcommand{\taxtis}{Ta$_{z}$Ti$_{1-z}$S$_2$}
\newcommand{\cotis}{Co$_{0.04}$TiS$_2$}
\newcommand{\bite}{Bi$_{2}$Te$_3$}

\newcommand{\ncm}[1]{\ensuremath{#1\,\times 10^{21} \textnormal{cm}^{-3}}}
\newcommand{\wkm}[1]{\ensuremath{#1\,\textnormal{W K}^{-1}\textnormal{m}^{-1}}}
\newcommand{\seebeck}[1]{\ensuremath{#1\,\mu\textnormal{V K}^{-1}}}

\begin{document}
\title{Intrinsic effects of substitution and intercalation on thermal transport in two-dimensional \tis{} single crystals.}

\author{Ramzy~Daou}
\affiliation{CRISMAT UMR6508, CNRS/Ensicaen/UCBN, 6 Boulevard du Mar\'echal Juin, F-14050 Caen Cedex 4, France}

\author{Hidefumi~Takahashi}
\affiliation{CRISMAT UMR6508, CNRS/Ensicaen/UCBN, 6 Boulevard du Mar\'echal Juin, F-14050 Caen Cedex 4, France}
\affiliation{Department of Physics, Nagoya University, Nagoya 464-8602, Japan}

\author{Sylvie~H\'ebert}
\affiliation{CRISMAT UMR6508, CNRS/Ensicaen/UCBN, 6 Boulevard du Mar\'echal Juin, F-14050 Caen Cedex 4, France}

\author{Marine~Beaumale}
\affiliation{CRISMAT UMR6508, CNRS/Ensicaen/UCBN, 6 Boulevard du Mar\'echal Juin, F-14050 Caen Cedex 4, France}

\author{Emmanuel~Guilmeau}
\affiliation{CRISMAT UMR6508, CNRS/Ensicaen/UCBN, 6 Boulevard du Mar\'echal Juin, F-14050 Caen Cedex 4, France}

\author{Antoine~Maignan}
\affiliation{CRISMAT UMR6508, CNRS/Ensicaen/UCBN, 6 Boulevard du Mar\'echal Juin, F-14050 Caen Cedex 4, France}

\date{\today}

\begin{abstract}

\textbf{The promising thermoelectric material \tis{} can be easily chemically doped and intercalated. We present here studies of single crystals that are intercalated with excess Ti or Co, or substituted with Ta. We demonstrate the intrinsic impact of these dopants on the thermal transport in the absence of grain boundary scattering. We show that Ta doping has the greatest impact on the thermal scattering rate per ion added, leading to a five-fold reduction in the lattice thermal conductivity as compared to stoichiometric single crystals.}

\end{abstract}
\maketitle

\section{Introduction}

Due to the rich variety of their physical properties, 2D materials are attracting much attention. Huge thermal conductivity values (\wkm{>10^3}) in graphene \cite{1} and in boron nitride \cite{2}, various electronic properties in transition metal dichalcogenides \cite{3}, high-T$_C$ superconductivity in cuprates \cite{4} are among some of the specific properties of interest. About two-thirds of the MX$_2$ transition metal dichalcogenides \cite{3} crystallize in layered structures and they exhibit a wide range of tunable electronic states (semiconducting, insulating, semi-metallic and superconducting), as well as exotic optical and mechanical properties. Their crystals can be easily cleaved as successive X layers are only weakly bonded by van der Waals forces.
 
The well-known sulfides derived from \tis{} have recently been under active investigation in the new context of thermoelectric materials research \cite{5,6,7,8}. 
While the thermoelectric power factor, $P_F  = S^2/\rho$, ($\rho$ and $S$ are the electrical resistivity and Seebeck coefficient, respectively) of \tis{} single crystals is similar to that of \bite{}, the thermoelectric material most commonly used in applications, the thermal conductivity ($\kappa$) exhibits relatively high values\cite{9}: $\kappa(300K)=\wkm{6.8}$, compared to \wkm{1.2} for \bite{} \cite{16}. This reduces the dimensionless figure of merit, $ZT = P_F T/\kappa$, which measures the theoretical efficiency of energy conversion. Several attempts have been made to reduce $\kappa$ in dense ceramics, for example via intercalation in the van der Waals gap as in Cu$_x$\tis{}, or substitution at either anion or cation sites as in TiS$_{2-y}$Se$_y$ or Ta$_z$Ti$_{1-z}$S$_2$ \cite{6,7,8}. These investigations have lead to ZT values of 0.4 at 650\,K.

However, for all these ceramic series, it is difficult to distinguish whether changes in $\kappa$ are a result of changes in the charge carrier concentration, the concentration of substituted or intercalated ions, or grain boundary effects. Moreover, the chemical formula should more properly be written as non-stoichiometric Ti$_{1+x}$S$_2$. It is known to be a degenerate extrinsic semiconductor with extra Ti$^{4+}$ cations in the van der Waals gap, the donated electrons of which populate the empty Ti d-band of the \tis{} layer \cite{10}. Meanwhile, the crystallographic texturation of ceramic samples implies that the transport properties always result from a convolution of in-plane and out-of-plane contributions, not to mention scattering by grain boundaries.

In this article, we report on in-plane electronic and thermal transport for several metallic single crystals derived from \tis. The changes in $\kappa$ (in-plane) are dominated by the lattice part with a peak at $\Theta_D$/5 ($\Theta_D$=Debye Temperature) expected for phonon-dominated heat transport. 
We show that reduction in the lattice heat transport ($\kappa_{lattice}$) deriving from the substitution or intercalation of foreign cations accounts for the majority of the change in $\kappa$, and we identify the species that are most effective at reducing the thermal conductivity.
The $\kappa_{lattice}$(300K) values obtained for crystals derived from \tis{} can be reduced by a factor of around five, from \wkm{6.3} (for the most stoichiometric crystals \cite{9}) to \wkm{1.2} (this study). We conclude that the thermoelectric properties of the 2D dichalcogenides could be improved by thermal conductivity engineering, even before nanostructuring is pursued.

\section{Methods}

Large platelet-like crystals 
were grown by a vapour transport technique using iodine as the transport agent \cite{11}. Intercalated and chemically substituted single crystals were grown by pre-reacting Co and Ta with \tis{} precursor at 500\,$^\circ$C prior to the growth phase. 
Shiny thin crystals were obtained for all the different batches and powder X-ray diffraction confirmed the presence of the \tis{} phase. 
Scanning electron microscopy with coupled EDX chemical analysis on all samples confirmed the nominal compositions within an uncertainty of $\pm 5$\%. In the case of \cotis{}, this means that Co was detected but with greater error than the nominal composition.
One ceramic sample of \tis{} grown with a slight nominal excess of Ti and densified by Spark Plasma Sintering was also used for comparison to the single crystals. Full details of this sample are in Ref.~\onlinecite{Beaumale2014}.

\begin{figure}
\includegraphics[width=0.47\textwidth]{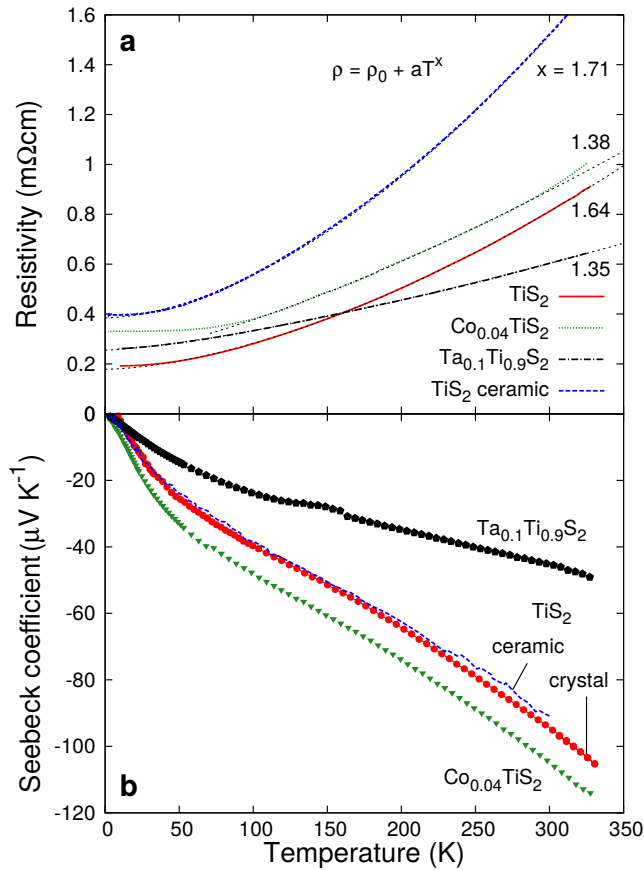} 
\caption{{\bf a)} Electrical resistivity of the \tis{} single crystals and derivatives, as well as the ceramic sample. The resistivity of the densified ceramic \tis{} is much higher than the single crystal despite very similar charge carrier concentration. Dashed lines show the fits to $\rho_0+ aT^x$. {\bf b)} Seebeck coefficient of the same samples.}
\label{fig:tep}
\end{figure}

Transport measurements were made on single crystal platelets of typical dimension $5 \times 1 \times 0.02$\,mm. Electrical contacts were made using Dupont 4929 silver paste.
Electrical resistivity and Hall effect was measured using a standard six probe technique. Thermal conductivity and Seebeck coefficient were measured on single crystals using a custom built steady-state experiment installed in a Physical Properties Measurement System cryostat. Calibrated heat pipes were used to measure the thermal power entering and leaving the sample \cite{13}. Power loss (principally from radiation) did not exceed 30\% at 300\,K. This contributes to an additional uncertainty in the thermal conductivity measurements of approximately 10\% at 300\,K, which becomes negligible below 150\,K. One crystal of \tis{} was also measured using the Harman technique \cite{15} and the results are in good agreement.

\section{Results and Discussion}
\subsection{Electrical Transport}

Electrical resistivity of the samples is shown in Fig.~\ref{fig:tep}a. 
They show metallic-like behavior with $\rho(300K)$ in between 0.6 and 1\,$\textnormal{m}\Omega \textnormal{cm}$ and resistivity ratio $\rho(300K)/\rho(5K) = 4.5$ related to residual resistivity values at 5\,K of 0.2 to 0.3\,$\textnormal{m}\Omega \textnormal{cm}$. The temperature dependence of the curves follows the previously seen $\rho = \rho_0 + aT^x$  with $1.3<x<2$ behaviour over a wide temperature range (60-300\,K) that has been ascribed to the approximately equal contributions of inter- and intra-valley scattering to the relaxation time \cite{9}. Values of $\rho_0$ and $x$ are shown in Table~\ref{tab:params} and the fits are included in Fig.~\ref{fig:tep}a. 

A key problem with \tis{} and derivatives is controlling the charge carrier concentration, which arises both from deliberate substitution or intercalation, and from the tendency for excess Ti$^{4+}$ to self-intercalate in the van der Waals gap. In the following, we establish the concentration of substituted and/or intercalated cations, $n_{ion}$, by reference to well-characterised ceramic series using the c-axis lattice parameters \cite{8,Beaumale2014,Inoue1989}, and obtain the corresponding charge carrier concentration, $n_e$, from Hall effect measurements.

In the case of the \tis{} single crystal, $c = 5.700 \AA$ gives\cite{Beaumale2014} $n_{ion} = \ncm{0.34(8)}$ and $n_e = \ncm{1.8(2)}$. These values are consistent with a stoichiometry close to \tixs{1.025}, assuming that all of the excess Ti$^{4+}$ is intercalated and that all four electrons become free carriers. The ceramic sample of the same nominal stoichiometry presents nearly identical Seebeck coefficient, which we consider to be a more reliable measure of the doping level in ceramic materials than $n_e$, as was shown in the cuprate superconductors \cite{Obertelli1992}. The value of $|S(300K)| = \seebeck{250}$ for the single crystals of Ref.~\onlinecite{9} corresponds to between 1 and 1.5\% excess titanitum according to the same ceramic series\cite{Beaumale2014}.

The \cotis{} crystal has $n_e = \ncm{1.1(2)}$ and a correspondingly larger $|S(300K)| = \seebeck{106}$. Previous studies of Co intercalation at high concentrations were consistent with the Co$^{2+}$ oxidation state \cite{22}. $c=5.683\AA$ implies a stoichiometry close to the nominal Co$_{0.04}$ composition\cite{Inoue1989}, and shows the contraction of the lattice characteristic of the strong covalent nature of bonding between the Co$^{2+}$ ion and the sulphur atoms in adjacent layers\cite{Kawasaki2011}. By contrast, the presence of intercalated Ti$^{4+}$ tends to enlarge the van der Waals gap.
The measured $n_e$ therefore corresponds to $n_{ion} = \ncm{0.55}$ and a stoichiometry Co$_{0.032}$; this is certainly within the margin of error afforded by the lattice parameter measurements. There is no suggestion of significant Ti self-intercalation in this case, and we speculate that the contraction of the lattice leads to a much lower ability of Ti$^{4+}$ to enter van der Waals gap.

The highly doped \tatis{} has a much smaller $|S(300K)| = \seebeck{40}$ and a correspondingly larger $n_e=\ncm{3.7(4)}$. Ta substitution at the Ti site provides one free carrier per ion, as can be seen from the linear dependence of $n_e$ on doping in Ref.~\onlinecite{8}. The value of $|S(300K)|$ is close to that obtained for a ceramic sample of the same nominal Ta$_{0.1}$ composition \cite{8}, as is $c=5.733\AA$. Given this composition, only \ncm{1.74} charge carriers are expected. We conclude that the rest of the carriers arise from excess Ti intercalation, such that the composition is approximately Ta$_{0.1}$Ti$_{0.93}$S$_2$, giving $n_{ion}^{Ta} = \ncm{1.74}$ and $n_{ion}^{Ti} = \ncm{0.5}$. The nominal and derived chemical formulae for all of the single crystals are summarised in Table~\ref{tab:params}.

\subsection{Thermal Transport}

\begin{figure}
\centering
 \includegraphics[width=0.47\textwidth]{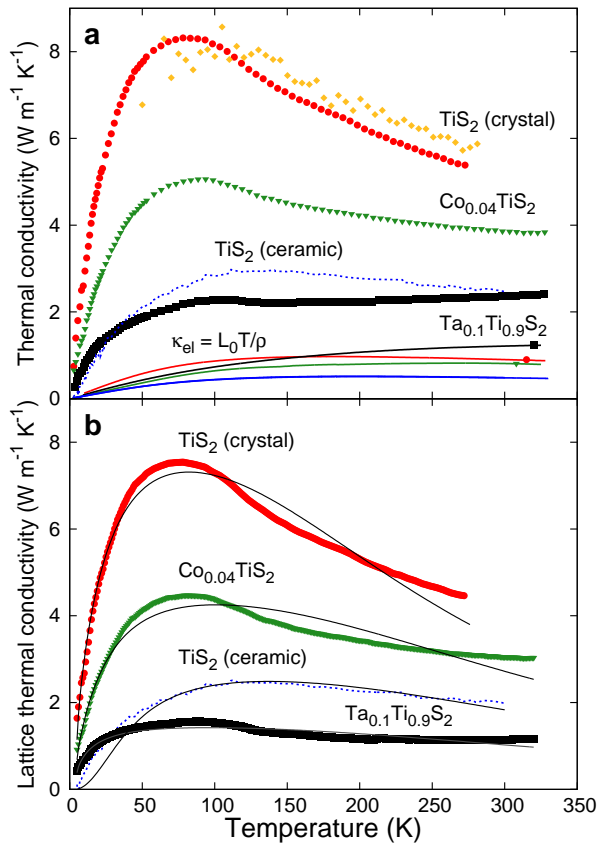}
\caption{{\bf a)} Thermal conductivity of single crystal compositions based on \tis{}. The red circles and yellow diamonds represent the steady state and Harman measurements on nominally stoichimetric \tis{} respectively (see text). The electronic contribution estimated from the Wiedemann-Franz law (solid lines) is subtracted to give the lattice thermal conductivity {\bf b)}. The peak height of $\kappa_{lattice}$ at $\sim 80$\,K is reduced as impurity scattering increases. The fits to the Debye-Callaway model (black solid lines) are dominated by the variation in this parameter. The estimated minimum lattice thermal conductivity is also shown.}
\label{fig:kappa}
\end{figure}

The in-plane thermal conductivity (Fig.~\ref{fig:kappa}a), particularly the lattice contribution (Fig.~\ref{fig:kappa}b), is a powerful probe of $n_{ion}$, especially in the absence of grain boundaries. The only previously reported thermal conductivity value for a slightly self-intercalated \tis{} crystal\cite{9} was $\kappa(300K)=\wkm{6.8}$, with $S=\seebeck{-251}$ corresponding to $n_e = \ncm{0.28}$. For the latter, the electronic part of $\kappa$ was only \wkm{0.43} against $\kappa_{el} \approx \wkm{1.0}$ for the present \tis{} and \cotis{} crystals (assuming that the Wiedemann-Franz law is valid in this regime for electronic heat transport). Generally speaking, the thermal conductivity curves follow the typical form expected for heat transported mainly by phonons: a large quasi-linear rise at low temperature as phonons are excited, a peak limited by cation and defect concentration, followed by a continuing decrease at higher temperatures as phonon-phonon umklapp scattering processes become dominant. 

We extract the lattice thermal conductivity (Fig.~\ref{fig:kappa}b) by subtracting the electrical contribution to the total thermal conductivity according to the Wiedemann-Franz law, $\kappa_{lattice} =  \kappa_{total} - \kappa_{el} =  \kappa_{total} -  \frac{L_0T}{\rho}$, where we use the Sommerfeld value of the Lorenz number $L_0 = 2.44\times 10^{-8}$\,$\textrm{W} \Omega^{-1} \textrm{K}^{-2}$. $\kappa_{el}$ extracted in this way is usually an upper bound since a) inelastic scattering processes, maximally active at $\Theta_D/5$, scatter the thermal but not the electrical current effectively and b) if the temperature broadening of the Fermi distribution approaches the bandwidth, scattering processes are no longer quasi-elastic. For the high carrier concentrations in the samples measured here, the Fermi temperature is in excess of 1000\,K and our measurements remain in the quasi-elastic regime \cite{10}. 

We analyse $\kappa_{lattice}$ using the Debye-Callaway model with contributions from boundary, impurity, umklapp and planar defect scattering:
\begin{equation}
 \kappa_{lattice} = \frac{k_B}{2\pi^2 v_{ph}} \int^{\Theta_D/T}_{0} \tau(x,T) \frac{x^4e^x}{(e^x -1)^2}
\label{eqn:dc}
\end{equation}
where $1/\tau = v_{ph}/d + A\omega^4 + B\omega^2T^3e^{\Theta_D/bT} + C\omega^2$ and $x = \hbar\omega/k_BT$. $v_{ph} = 3793$\,$\textnormal{ms}^{-1}$ is the in-plane sound velocity \cite{17} and $d=1$\,mm is the relevant limit for boundary scattering, corresponding to the average sample dimension. $A$ is the prefactor for scattering by point-like defects (lattice imperfections or foreign cations), $B$ and $b$ are the parameters describing the umklapp processes, and $C$ is related to the density of planar defects. The inclusion of $C$ significantly improves the quality of the fits at low temperature. The impurity, defect and boundary scattering control the shape of the curve at low temperatures, while umklapp scattering controls the decline at high temperatures. $\Theta_D = 240$\,K is used for all samples\cite{Inoue1986}.
In the case of the ceramic sample, we use $d = 33$\,nm in order to compare to the results of Ref.~\onlinecite{8}. 

\begin{table*}[t]
 \begin{tabular}{ c | c  c  c  c  c  c | c  c  c  c}
  \hline
  Sample & c (\AA) & $n_e$(300K) & n$_{imp}$ & $\rho_0$ & $x$  & S(300K)  & A & B & b & C \\
   &  & ($10^{21} \textnormal{cm}^{-3}$) & ($10^{21} \textnormal{cm}^{-3}$) &(m$\Omega$cm) & $\rho\sim T^x$ & ($\mu$V K$^{-1}$) & ($K^{-4}s^{-1}$)& ($K^{-5}s^{-1}$) &  & ($10^6 K^{-2}s^{-1}$)\\
  \hline
  \tis{} (crystal) & 5.700 & 1.8(2) & 0.43 & 0.18 & 1.64  & -95  & 1200  & 0.374 & 4.6 & 3.61 \\
  \cotis{}         & 5.683 & 1.1(2) & 0.55 & 0.33 & 1.38  & -106 & 1850  & 0.153 & 4.5 & 7.21 \\
  \tatis{}         & 5.733 & 3.7(4) & 1.74(Ta), 0.5(Ti) & 0.25 & 1.35  & -45  & 13400 & 0.367 & 4.3 & 11.6\\
  \tis{} (ceramic) & 5.701 & 1.1(1) & 1.74 & 0.38 & 1.71 & -93  & 730  & 0.455 & 2.8 & 3.26 \\
  \hline
 \end{tabular}
\caption{Electronic properties and fit parameters of Debye-Callaway model of the lattice thermal conductivity of \tis{} and derivatives discussed in the text.}
\label{tab:params}
\end{table*}

The results (fits are shown in Fig.~\ref{fig:kappa}b and parameters in Table~\ref{tab:params}) show that the impurity and defect scattering parameters ($A$ and $C$) for the single crystals trend as expected, increasing as the peak in thermal conductivity around 80\,K is reduced. 
The umklapp scattering parameters do not show a discernible trend but are all of similar magnitude; these are strongly influenced by the high temperature behaviour where uncertainty in the measurements leads to rather large uncertainty in the extracted parameters. Additionally, the single crystal data all have a similar high-temperature form, with a slight concavity near 130\,K. This suggests that the variation in $\kappa_{lattice}$ is dominated by the changes in the foreign cation concentration, as is supported by the fit parameters. A similar trend was observed in a series of \taxtis{} ceramics\cite{8}.

The $A$ parameter as modelled by Klemens\cite{21} is proportional to three contributions, as shown in Eqn.~\ref{eqn:impurity}:
\begin{equation}
 A \sim 
 \frac{n_{ion}}{n}
 \biggl[
 \biggl(\frac{\Delta M}{M}\biggr)^2 + \varepsilon\biggl(\frac{\Delta \delta}{\delta}\biggr)^2
 \biggr]
\label{eqn:impurity}
\end{equation}
where $n_{ion}/n$ is the fractional concentration of substitutions/intercalants. $\Delta M$ represents the difference in the mass of the foreign cation relative to the mass of the replaced ion, $M$. $\varepsilon$ is a parameter related to stiffness and anharmonicity that scales the influence of the size difference $\Delta \delta/\delta$ of the cation relative to the undisturbed site.

Given the proportionality to $n_{ion}$, we can thus calculate the effective $A$ parameter per  ion, $A_{ion}=A/n_{ion}$. In the case of two types of cation $A = \sum_i A_i$, assuming Mathiessen's rule holds, since different scattering rates are additive in Eqn~\ref{eqn:dc}. These are shown in Table~\ref{tab:As}. $A_{ion}$ for Ta$^{4+}$ is approximately 3 times the value for Ti$^{4+}$ intercalants, with Co$^{2+}$ intercalants being only slightly more effective scatterers than Ti$^{4+}$. The parameters $(\Delta M/M)^2$ and $(\Delta \delta/\delta)^2$ are estimated from the atomic masses and observed unit cell volume changes (Table~\ref{tab:As}). The very large mass of Ta contributes significantly, but it causes a relatively small change in unit cell volume. By contrast, Ti and Co intercalants have a smaller mass effect. For these, we have assumed $M=37.3$ atomic mass units, the unit cell average, in order to try to make a direct comparison with the Ta case, but this form of Eqn.~\ref{eqn:impurity} is not well suited to intercalants, where a zero mass void is replaced by an ion. The volume changes per ion are larger for Ti and Co than for Ta, and so the associated contributions to the scattering rate are not as small as the mass parameters would suggest.

\begin{table}
 \begin{tabular}{ c | c  c  c }
  \hline
   ion       & $A_{ion}$ ($10^{-21} \textnormal{cm}^{3}K^{-4}s^{-1}$) & $(\Delta M/M)^2$ & $(\Delta \delta/\delta)^2$\\
  \hline
  Ti$^{4+}$  & 2670 & 0.082 & 0.31  \\
  Co$^{2+}$  & 3360 & 0.33  & 0.15  \\
  Ta$^{4+}$  & 6900 & 8.6   & 0.034 \\
  \hline
 \end{tabular}
\caption{Effective parameters for scattering rate per substituted/intercalated cation in \tis{} single crystals.}
\label{tab:As}
\end{table}

The ceramic sample shows a smaller $A$ than the single crystal of \tis{}, but also a much lower $\kappa_{lattice}$. The primary reason for the difference between these two cases is the parameter $d$, which describes scattering by grain boundaries. If we take the identical parameters as for the single crystal, a value of $d = 60$\,nm produces a very similar $\kappa_{lattice}$ to the one shown in Fig.~\ref{fig:kappa}b, showing that this is indeed the dominant scattering mechanism. The reduction in thermal conductivity on going from crystal to ceramic is almost perfectly balanced by the increase in resistivity, however, leading to no change in $ZT$ at 300\,K.

\section{Conclusions}

In conclusion, we have shown using well calibrated concentrations of substituted or intercalated ions in single crystal materials that the most effective species at reducing the thermal conductivity in the \tis{} system is Ta$^{4+}$ substitution at the Ti site. Ta substitution is particularly suitable as it provides only 1 electron per ion, compared to 2 and 4 for Co and Ti intercalants, introducing more scattering centers for a given charge carrier concentration. The thermoelectric performance of Ta doped materials so far studied has however been limited by parasitic self-intercalation of Ti. Improvements in thermoelectric performance will only be realised once this effect can be controlled. Then studies of ceramic \tis{} and derivatives for potential applications can be pursued, as similar performance can be obtained even in the presence of grain boundaries.

\end{document}